\documentclass[prl,twocolumn,showpacs,preprintnumbers,amsmath,amssymb]{revtex4}
\usepackage{epsfig}
\usepackage{amsmath,amssymb}
\usepackage{graphicx}
\usepackage{dcolumn}
\usepackage{bm}
\usepackage{mathrsfs}

\begin{document}

\title{Large Scale Quantum Computation in an Anharmonic Linear Ion Trap}
\author{G.-D. Lin$^{1}$, S.-L. Zhu$^{2,1}$, R. Islam$^{3}$, K. Kim$^{3}$, M.-S. Chang$^{3}$, S. Korenblit$^{3}$, C. Monroe$^{3}$,
and L.-M. Duan$^{1}$}
\affiliation{$^{1}$FOCUS Center and MCTP,
Department of Physics, University of Michigan,
Ann Arbor, Michigan 48109\\
$^{2}$LQT and ICMP, Department of Physics, South China Normal University, Guangzhou, China\\
$^{3}$JQI and Department of Physics, University of Maryland, College
Park, MD 20742}
\date{\today }

\begin{abstract}
We propose a large-scale quantum computer architecture by stabilizing a
single large linear ion chain in a very simple trap geometry. By confining
ions in an anharmonic linear trap with nearly uniform spacing between ions,
we show that high-fidelity quantum gates can be realized in large linear ion
crystals under the Doppler temperature based on coupling to a near-continuum
of transverse motional modes with simple shaped laser pulses.
\end{abstract}


\maketitle


Trapped atomic ions remain one of the most attractive candidates for
the realization of a quantum computer, owing to their long-lived
internal qubit coherence and strong laser-mediated Coulomb
interaction \cite{1,2,3,4}. Various quantum gate protocols have been
proposed \cite{1,10,11,12,13,14} and many have been demonstrated
with small numbers of ions \cite{4,5,6,7,8,16}. The central
challenge now is to scale up the number of trapped ion qubits to a
level where the quantum behavior of the system cannot be efficiently
modeled through classical means \cite{4}. The linear rf (Paul) trap
has been the workhorse for ion trap quantum computing, with atomic
ions laser-cooled and confined in 1D crystals \cite{1,2,3,4}
(although there are proposals for the use of 2D crystals in a
Penning trap \cite{17} or array of microtraps \cite{11}). However,
scaling the linear ion
trap to interesting numbers of ions poses significant difficulties \cite{2,4}%
. As more ions are added to a harmonic axial potential, a structural
instability causes the linear chain to buckle near the middle into a zigzag
shape \cite{18}, and the resulting low-frequency transverse modes and the
off-axis rf micromotion of the ions makes gate operation unreliable and
noisy. Even in a linear chain, the complex motional mode spectrum of many
ions makes it difficult to resolve individual modes for quantum gate
operations, and to sufficiently laser cool many low-frequency modes. One
promising approach is to operate with small linear ion chains and multiplex
the system by shuttling ions between multiple chains through a maze of
trapping zones, but this requires complicated electrode structures and
exquisite control of ion trajectories \cite{2,9}.

In this paper, we propose a new approach to ion quantum computation in a
large linear architecture that circumvents the above difficulties. This
scheme is based on combination of several ideas. First, an anharmonic axial
trap provided by static electrode potentials can stabilize a single linear
crystal containing a large number of ions. Second, tightly-confined and
closely-spaced transverse phonon modes can mediate quantum gate operations
in a large architecture \cite{19}, while eliminating the need for
single-mode resolution and multimode sideband cooling. Third, gate
operations on the large ion array exploit the local character of the
laser-induced dipole interaction that is dominated by nearby ions only. As a
result, the complexity of the quantum gate does not increase with the size
of the system.

\begin{figure}[bp]
\includegraphics[width=8cm]{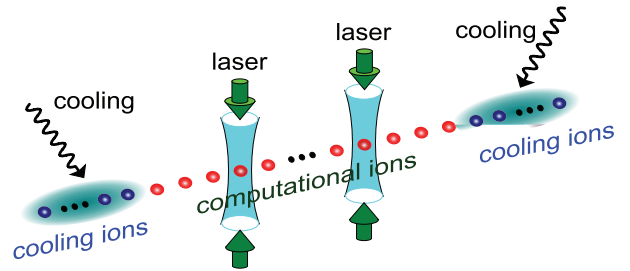}
\par
\caption{(Color online) (a) Linear architecture for large scale quantum
computation, where lasers address individual ions and couple to local modes
of the ions, while the edge ions are continuously Doppler laser-cooled.}
\label{fig:archi}
\end{figure}

The proposed ion architecture is illustrated in Fig. \ref{fig:archi}. It is a large linear
array where the strong confinement in the transverse $(x,y)$ direction is
provided by the ponderomotive Paul trap with an effective potential of the
form $V(x,y)=\left( m\omega _{x}^{2}x^{2}+m\omega _{y}^{2}y^{2}\right) /2$,
where $m$ is the mass of each ion. The ions are initially Doppler cooled,
with a number of ions at the edges of the chain continuously Doppler cooled
in order to overwhelm any heating that occurs during the gate operation. The
middle portion and majority of the ion chain is used for quantum
computation. Given an appropriate axial static potential $V(z)$ from the
axially-segmented electrodes, we assume these computational ions are
distributed nearly uniformly, with a neighboring distance of about $\sim 10$
$\mu m$. This enables efficient spatial addressing with focused laser beams
along the transverse direction for quantum gate operations.

When the axial potential takes the conventional harmonic form
$V(z)=m\omega _{z}^{2}z^{2}/2$, the ion array is subject to the
well-known zigzag transition unless the trap anisotropy is at least
$\omega _{x,y}/\omega _{z}>0.77N/\sqrt{\log N}$ \cite{2,20}, where
$N$ is the number of ions. This structural instability is caused by
the spatial inhomogeneousity of the ion distribution. When the ions
are instead uniformly spaced by neighboring distance $d_{0}$, it is
easy to see that the linear structure is always stable even for an
infinite chain so long as $\omega _{x,y}^{2}>7\zeta
(3)e^{2}/(2md_{0}^{3})\approx 4.2e^{2}/(md_{0}^{3})$, where $\zeta
(l)$ is the Reimann Zeta function and $e$ is the charge of each ion.
Therefore, a large linear structure can be stabilized so long as
appropriate static potentials from the trap electrodes to make the
ion distribution homogeneous along the axis.

\begin{figure}[bp]
\includegraphics[width=8cm]{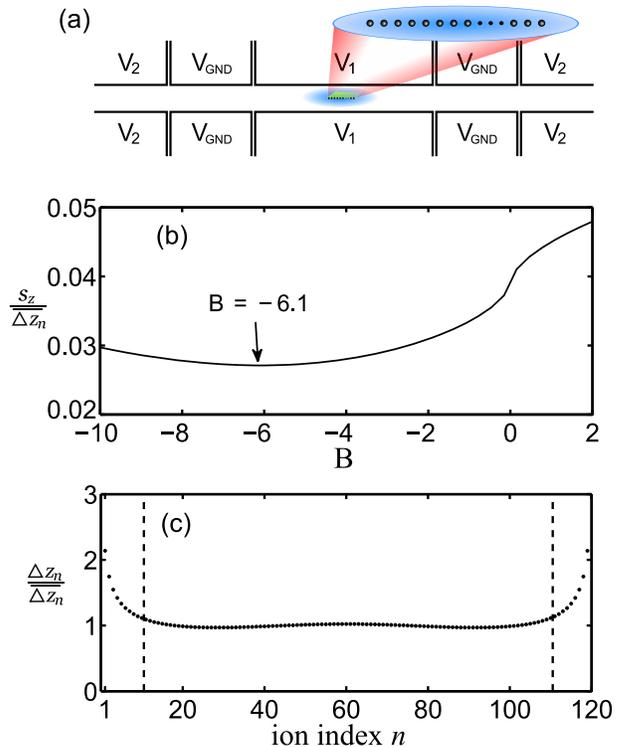}
\par
\caption{(Color online) (a) Sample five-segment linear ion trap with
voltages $V_i (i=1,2)$ to produce a quartic axial potential. The ions are
confined in the central segment. (b) The variance of the ion spacings $s_z$
in a linear quartic trap as a function of the trap parameter $B$ that
characterizes the ratio of quadratic to positive-quartic nature of the
potential. (c) The distribution of the ion spacings at the optimum value
$B=-6.1$. The computational ions are within the dashed lines, where the
spacing is essentially uniform.}
\label{fig:homogeneity}
\end{figure}

To illustrate the general method, here we consider an explicit
example with a quartic potential $V(z)=\alpha _{2}z^{2}/2+\alpha
_{4}z^{4}/4$ that can be realized with a simple five-segment
electrode geometry as shown in Fig. \ref{fig:homogeneity}a. Under a
quartic trap $V(z)$, the axial equilibrium position $z_{i}$ of the
$i $th ion can be obtained by solving the force balance equations
$\partial U/\partial z_{i}=0$, where $U=\sum_{i}\left[
V(z_{i})+V(x,y)\right] +\sum_{i<j}e^{2}/\left\vert
\mathbf{r}_{i}-\mathbf{r}_{j}\right\vert $ is the overall potential
including the ions' mutual interactions. We optimize the
dimensionless ratio $B=|\alpha _{2}/e^{2}|^{2/3}(\alpha _{2}/\alpha
_{4}) $ characterizing the axial potential to produce a
nearly-uniformly spaced crystal. To be concrete, we consider an
array of $120$ ions, with $10$ ions at each edge continuously laser
cooled and $100$ qubit ions in the middle for coherent quantum gate
operation. We solve the equilibrium positions of all the ions under
$V(z)$ (see appendix A) and minimize the variance in ion
spacing $s_{z}=\sqrt{\frac{1}{100}\sum_{i=11}^{110}\left( \Delta z_{n}-%
\overline{\Delta z_{n}}\right) ^{2}}$ for the qubit ions, where $\Delta z_{n}
$ is the distance between the $n$th and ($n+1$)th ion in the chain and $%
\overline{\Delta z_{n}}$ denotes its average. The variance in spacing is
shown in Fig. \ref{fig:homogeneity}b as a function of the parameter $B$. The value of $s_{z}$ is
fairly insensitive to $B$ and reaches a minimum when $B\approx -6.1$. Here,
the distribution of ion spacings $z_{n}$ is shown in Fig. \ref{fig:homogeneity}c, which is
remarkably homogeneous for the qubit ions even though we have optimized just
one control parameter: $s_{z}/\overline{\Delta z_{n}}$ deviates by only $3\%$
over the entire crystal. In this configuration, if we take $\overline{\Delta
z_{n}}=10$ $\mu m$ for atomic Yb$^{+}$ ions, we only need a transverse
center-of-mass frequency $\omega _{x,y}/2\pi >221$ kHz to stabilize the
linear structure. In this paper, we actually take $\omega _{x}/2\pi =5$ MHz,
as is typical in experiments, and such transverse confinement would be able
to stabilize linear chains with thousands of ions under an optimized quartic
potential.

We now describe quantum gate operations with this large ion chain,
mediated by many transverse phonon modes. Given the equilibrium
positions of the ions, we can efficiently determine all axial and
transverse phonon modes. We then apply a spin-dependent laser force,
with the resulting interaction Hamiltonian \cite{21,22}
\begin{equation}
H=\sum_{n}\hbar \Omega _{n}(t)\sigma _{n}^{z}\cos (kq_{n}+\mu t),
\end{equation}%
where the transverse displacement $q_{n}$ of the $n$th ion in the $x$
direction is expressed in terms of phonon modes $a_{k}$ with eigenfrequency $%
\omega _{k}$ and the normal mode matrix $b_{n}^{k}$ by $q_{n}=%
\sum_{k}b_{n}^{k}\sqrt{\hbar /2m\omega _{k}}(a_{k}^{\dagger }e^{i\omega
_{k}t}+a_{k}e^{-i\omega _{k}t})$. The normal mode matrix $b_{n}^{k}$ and its
eigenfrequency $\omega _{k}$ are determined by solving the eigen-equations $%
\sum_{n}A_{in}b_{n}^{k}=\omega _{k}^{2}b_{i}^{k}$, where $A_{in}\equiv
\partial ^{2}U/\partial x_{i}\partial x_{n}$ are calculated at the ions'
equilibrium positions $z_{i}$. In Eq. (1), $\sigma _{n}^{z}$ is the Pauli
spin operator for the $n$th ion, $\Omega _{n}(t)$ denotes the Rabi frequency
of the laser pulse on the $n$th ion with detuning $\mu $ from the qubit
resonance, and the effective laser momentum kick $k$ is assumed to be along
the transverse $x$ direction. (For twin-beam stimulated Raman laser forces
and hyperfine state qubits, the effective laser kick is simply along the
difference wavevector $\mathbf{k_{1}}-\mathbf{k_{2}}$ of the two beams.) Due
to the strong transverse confinement, the Lamb-Dicke parameter $\eta
_{k}\equiv \left\vert k\right\vert \sqrt{\hbar /2m\omega _{k}}\ll 1$, and
the Hamiltonian $H$ can be expanded as $H=-\sum_{n,k}\hbar \chi
_{n}(t)g_{n}^{k}(a_{k}^{\dagger }e^{i\omega _{k}t}+a_{k}e^{-i\omega
_{k}t})\sigma _{n}^{z}$ with $g_{n}^{k}=\eta _{k}b_{n}^{k}$ and $\chi
_{n}(t)=\Omega _{n}(t)\sin (\mu t)$ (the effect of higher-order terms in the
Lamb-Dicke expansion will be estimated later). The corresponding evolution
operator is given by \cite{22}
\begin{eqnarray}
U(\tau)=\exp\left[i\sum_{n,k}[\alpha_{n}^{k}(\tau)a_{k}^{\dagger}+\alpha_{n}^{k\ast}(\tau)a_{k}]\sigma_{n}^{z}\right.\nonumber\\
 \left.+i\sum_{m<n}\phi_{mn}(\tau)\sigma_{m}^{z}\sigma_{n}^{z}\right],
\end{eqnarray}
where $\alpha _{n}^{k}(\tau )=\int_{0}^{\tau }\chi
_{n}(t)g_{n}^{k}e^{i\omega _{k}t}dt$ characterizes the residual entanglement
between ion $n$ and phonon mode $k$ and $\phi _{mn}(\tau )=2\int_{0}^{\tau
}dt_{2}\int_{0}^{t_{2}}dt_{1}\sum_{k}g_{m}^{k}g_{n}^{k}\chi _{m}(t_{2})\chi
_{n}(t_{1})\sin \omega _{k}(t_{2}-t_{1})$ represents the effective
qubit-qubit interaction between ions $m$ and $n$.

\begin{figure}[bp]
\includegraphics[width=8cm]{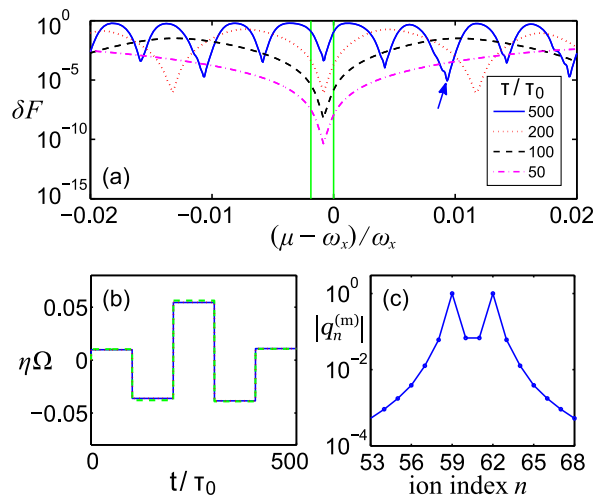}
\caption{(Color online) (a) The gate infidelity $\protect\delta F$ as a
function of the the laser detuning $\protect\mu$ from the qubit resonance,
with optimized Rabi frequencies over $M=5$ equal segments of the laser
pulse. Different curves correspond to different gate times. The $120$
transverse phonon modes are all distributed within the narrow frequency
range indicated by the two vertical lines. (b) The shape of the laser pulse
that achieves the optimal fidelity (with $\protect\delta F=8.5 \times
10^{-6} $), at the detuning shown by the arrow in Fig. 3a (with $(\protect\mu%
-\protect\omega_x)/\protect\omega_x=9.3 \times 10^{-3}$) and the gate time $%
\protect\tau=500 \protect\tau_0$. The dashed (or dotted) lines represent the
approximate optimal solutions of the laser shape where only $4$ (or $8$)
ions (from the 59th to 62th or the 57th to 64th, respectively) are allowed
to vibrate and all the other ions are fixed in their equilibrium positions.
This approximation doesn't significantly change the optimal laser pulse
shape compared to that of the exact solution (represented by the solid line)
where all ions are allowed to vibrate, so the gate is essentially local and
the gate complexity does not depend on the crystal size. (c) The relative
response of the ions for the gate shown in Fig. \ref{fig:fg}b (characterized by the
largest spin-dependent shift $\left\vert q_n^{(m)}\right\vert$ during the
gate time $\protect\tau$). We take a relative unit where $\left\vert
q_n^{(m)}\right\vert$ for the target ions haven been normalized to $1$. The
fast decay of the response as one moves away from the target ions (59th ad
62th) shows that the gate involves vibration of only local ions.}
\label{fig:fg}
\end{figure}

For a two-qubit gate on an ion pair $i$ and $j$, we direct laser light
exclusively on these two ions ($\Omega _{i}(t)=\Omega _{j}(t)\equiv \Omega
(t)$ and all other $\Omega _{n}(t)=0$), and the evolution operator reduces
to the standard controlled $\pi $-phase (CP) gate for $\alpha _{n}^{k}(\tau
)=0$ and $\phi _{jn}(\tau )=\pi /4$. For a large ion crystal, the residual
entanglement with the motional modes cannot be eliminated completely, but we
can minimize the resulting gate infidelity by optimizing the laser pulse
shape $\Omega (t)$ \cite{22}. Assuming each phonon mode $k$ is cooled to
temperature $T_{k}$, the infidelity of the CP gate from the residual
motional entanglement is given by $\delta F=[6-2(\Gamma _{i}+\Gamma
_{j})-\Gamma _{+}-\Gamma _{-}]/8$ \cite{22}, where $\Gamma _{i(j)}=\exp
[-\sum_{k}|\alpha _{i(j)}^{k}(\tau )|^{2}\bar{\beta}_{k}/2]$, $\Gamma _{\pm
}=\exp [-\sum_{k}|\alpha _{i}^{k}(\tau )\pm \alpha _{j}^{k}(\tau )|^{2}\bar{%
\beta}_{k}/2]$, and $\bar{\beta}_{k}=\coth (\hbar \omega _{k}/k_{B}T_{k})$.

To minimize the gate infidelity $\delta F$, we break the laser pulse on the
two ions into uniform segments of constant intensity as shown in Fig. \ref{fig:fg}b and
optimize the values $\Omega ^{(i)}$ $\left( i=1,\ldots ,M\right) $ over $M$
equal-time segments \cite{22}. The control of $\Omega \left( t\right) $ from
one segment to the next can easily be accomplished with optical modulators.
(Alternatively, we can modulate the detuning $\mu $ of the laser pulse as a
control parameter.) After optimization of $\Omega ^{\left( i\right) }$, the
infidelity $\delta F$ is shown as a function of the detuning $\mu $ in Fig.
\ref{fig:fg}a for $M=5$ segments. For this example, we perform a CP gate on the $59$th
and $62$ th ions in this $120$-ion chain. With an appropriate choice of $%
\Omega ^{\left( i\right) }$ and $\mu $, the infidelity can be made
negligible (well below $10^{-5}$). For this calculation, we assume Doppler
cooling for all modes and take gate times $\tau $ in the range $50\tau _{0}$
to $500\tau _{0}$ where $\tau _{0}=2\pi /\omega _{x}=0.2\mu s$ is the period
of transverse harmonic motion. The gate can certainly be faster with
stronger laser beams (there is no speed limit), and with a faster gate, the
control becomes easier as the gate becomes more localized (Fig. \ref{fig:fg}).

Interestingly, we use only a few control parameters ($M=5$ segments) to
perform a high-fidelity gate that involves excitation of hundreds of
transverse phonon normal modes. This is possible because the gate has a
local character where the contribution to the CP gate comes primarily from
the spin-dependent oscillations of the ions that are close to the target
ions. To show this, we plot the response of each ion in Fig. \ref{fig:fg}c during the
gate operation. Note that the displacement $q_{n}$ of the $n$th ion is
spin-dependent during the gate, and we can use its largest magnitude $%
\left\vert q_n^{(m)}\right\vert$ over the gate time $\tau $ to characterize
the response of ion $n$, as is shown in Fig. \ref{fig:fg}c. The ion response decays
very fast from the target ions ($59$th and $62$th in this case) and can be
safely neglected after a distance of a few ions. Thus during a gate, only
the motion of ions near the target ions are important, and the other ions
largely remain in their equilibrium positions. The resultant control
parameters from this approximation are almost identical to those shown in
Fig. \ref{fig:fg}b. Owing to the local character of the gate, the complexity of a gate
operation does not depend on the chain size, and we can perform gates in
parallel on ions in different regions of a large chain.

We now discuss several sources of noise for gates in a large ion crystal and
show that their effects are negligible. First, the axial ion modes have
large phonon occupation numbers under Doppler cooling alone, and the
resulting thermal spread in position along the axial direction can degrade
the effective laser interaction. For example, the lowest axial mode in a $120
$-ion chain of Yb$^{+}$ ions with a spacing $\overline{z}\sim 10\mu $m has a
frequency of only $\omega _{L0}/2\pi =9.8$ kHz and a mean thermal phonon
number $\overline{n}_{0}\approx \gamma /\omega _{L0}\approx 10^{3}$ under
Doppler laser cooling (radiative half-linewidth $\gamma /2\pi =10$ MHz). We
assume the quantum gate laser beams are directed along the transverse
direction with an axial Gaussian laser profile $\Omega \left( z\right)
\propto e^{-(z/w)^{2}}$ centered on each ion. The beam waist is taken as $w=%
\overline{z}/2.5\approx 4\mu m$ so that the cross-talk error probability
between adjacent ions is $P_{c}=e^{-2(\overline{z}/w)^{2}}<10^{-5}$. The
position fluctuation $\delta z_{n}$ of the $n$th ion causes the effective
Rabi frequency to fluctuate, resulting in a gate infidelity $\delta
F_{1}\approx \left( \pi ^{2}/4\right) \left( \delta \Omega _{n}/\bar{\Omega}%
_{n}\right) ^{2}\approx \left( \pi ^{2}/4\right) \left( \delta
z_{n}/w\right) ^{4}$. The fluctuation $\delta z_{n}$ can be calculated
exactly from summation of contributions of all the axial modes and its value
is almost independent of the index $n$ for the computational ions (see
Appendix). Under Doppler laser-cooling, $\overline{\delta z_{n}}\approx
0.26\mu m$ and the corresponding infidelity is $\delta F_{1}=4.4\times
10^{-5}$. The position fluctuation of the ions may also lead to anharmonic
ion motion, whose contribution to the gate infidelity can be estimated by $%
\delta F_{2}\sim \left( \delta z_{n}/\overline{z}\right) ^{2}\sim 6.8\times
10^{-4}$. Finally, in the transverse direction we estimate the infidelity
caused by the higher-order expansions in the Lamb-Dicke parameter. As all
the transverse modes have roughly the same frequency $\omega _{k}\approx
\omega _{x}$, the effective Lamb-Dicke parameter for the transverse modes is
$\eta _{x}=\left\vert \Delta \mathbf{k}\right\vert \sqrt{\hbar /2m\omega _{x}%
}\approx 0.038$ for Yb$^{+}$ ions at $\omega _{x}/2\pi =5$ MHz, with each
mode containing a mean thermal phonon number $\bar{n}_{x}\approx 2.0$ under
Doppler cooling. The resultant gate infidelity is estimated to be $\delta
F_{3}\approx \pi ^{2}\eta _{x}^{4}\left( \bar{n}_{x}^{2}+\bar{n}%
_{x}+1/8\right) \approx 7\times 10^{-4}$ \cite{19,23}. Note finally that
sideband cooling is possible in the transverse direction as all the modes
have nearly the same frequency, thus reducing the gate infidelity due to
transverse thermal motion by another order of magnitude.

In summary, we have shown through explicit examples and calculations that it
is feasible to stabilize large linear ion crystals where the gate complexity
does not increase with the size of the crystal and the gate infidelity from
thermal fluctuations can be made negligibly small under routine Doppler
cooling. The results suggest a realistic prospect for realization of large
scale quantum computation in a simple linear ion architecture.

\textbf{Appendix: Thermal position fluctuation of the ions along the axial
direction}

We address individual ions through focused laser beams which typically take
a Gaussian shape along the $z$ direction with $\Omega _{n}(z)\propto
e^{-z_{n}^{\prime 2}/\Delta ^{2}}$, where $z_{n}^{\prime }=z-z_{n}$ is
centered at the equilibrium position $z_{n}$ of the $n$th ion. Under the
Doppler temperature, the ions have significant thermal fluctuation of their
positions along the $z$ direction, which leads to an effectively fluctuating
laser amplitude $\Omega _{n}(z)$ and induces infidelity to the gate
operation. This position fluctuation influences both the single-bit and the
two-bit operations in the same way. To quantify the gate error caused by
this fluctuation, let us consider a spin-flip gate operated on the $n$th ion
as a typical example. For a spin-flip with a $\pi -$pulse, the gate fidelity
is given by $F_{1}=\sin ^{2}(\bar{\Omega}_{n}\tau +\delta \Omega _{n}\tau
)\approx 1-(\pi ^{2}/4)(\delta \Omega _{n}/\bar{\Omega}_{n})^{2}$, where $%
\bar{\Omega}_{n}$ is the expectation value of the Rabi frequency which
satisfies $\bar{\Omega}_{n}\tau =\pi /2$ for a spin-flip gate, and $\delta
\Omega _{n}$ is its fluctuation caused by the position fluctuation of the
ion. From $\Omega _{n}(z)\propto e^{-z_{n}^{\prime 2}/\Delta ^{2}}\approx
1-z_{n}^{\prime 2}/\Delta ^{2}$ around the equilibrium position, the gate
infidelity $\delta F_{1}\equiv 1-F_{1}=(\pi ^{2}/4)\left( \delta
z_{n}/\Delta \right) ^{4}$, where $\delta z_{n}\equiv \left( \overline{%
z_{n}^{\prime 4}}-\overline{z_{n}^{\prime 2}}^{2}\right) ^{1/4}$
characterizes the thermal position fluctuation of the $n$th ion along the
axial direction.

\begin{figure}[bp]
\includegraphics[width=8cm]{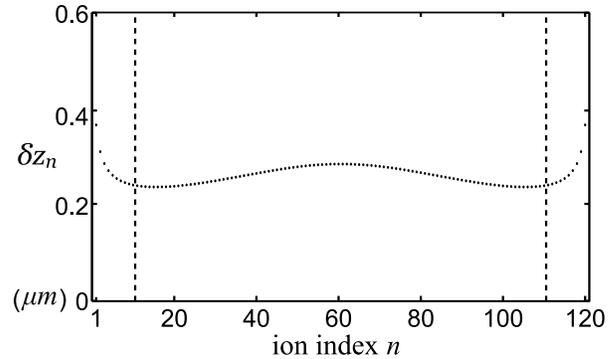}
\caption{The axial position fluctuation $\delta z_{n}$ is plotted
along the ion chain, which is about $0.26\mu\mbox{m}$ for the computational
ions ($n$ from $11$ to $110$).}
\label{fig:fluctuation}
\end{figure}

The phonon modes are in thermal equilibrium under the Doppler temperature $T$
, with their density operator given by $\rho
_{m}=\prod_{k}\sum_{\{n_{k}\}}P_{k}\left\vert n_{k}\right\rangle
\left\langle n_{k}\right\vert $, where $P_{k}=\bar{n}_{k}^{n_{k}}/\left(
\bar{n}_{k}+1\right) ^{n_{k}+1}$ is the probability of having $n_{k}$
phonons in the $k$th mode, and $\bar{n}_{k}=k_{B}T/(\hbar \omega _{k})$ is
the average phonon number. From $\overline{z_{n}^{\prime 2}}=\text{tr}%
(z_{n}^{\prime 2}\rho _{m})$ and $\overline{z_{n}^{\prime 4}}=\text{tr}%
(z_{n}^{\prime 4}\rho _{m})$, we explicitly have $\delta z_{n}=\sqrt{\hbar
/2m}\left[ \sqrt{2}\sum_{k}\left( b_{n}^{z,k}\right) ^{2}\left( 2\bar{n}%
_{k}+1\right) /\omega _{z,k}\right] ^{1/2}$where $\omega _{z,k}$ and $%
b_{n}^{z,k}$ denote th eigen-frequencies and eigen-matrices of the axial
modes. For our example of a $120$-ion chain with the Doppler temperature $%
k_{B}T/\hbar =62\mbox{MHz}$ for the Yb$^{+}$ ions, the resultant position
fluctuation $\delta z_{n}$ is plotted in Fig. \ref{fig:fluctuation} for all
the ions. One can see that for the computational ions (n from 11 to 110), $%
\delta z_{n}\approx 0.26$ $\mu m$ with its value almost independent of the
ion index. The position fluctuation is still significantly smaller than the
ion spacing ($\approx 10$ $\mu m$), which ensures a tiny gate infidelity $%
\delta F_{1}$ as discussed in the main text.

\acknowledgements

This work is supported by IARPA under ARO contract
W911NF-04-1-0234W911NF-08-1-0355, the DARPA OLE Program under ARO
Award W911NF-07-1-0576, and the MURI Quantum Simulation Program
under AFOSR contract. S.L.Z. is supported by the NSF and SKPBR of
China.

\end{document}